\documentclass[10pt]{iopart}
\usepackage{iopams}
\usepackage{xcolor,graphicx,ulem}
\usepackage[colorlinks=true,urlcolor=blue,citecolor=blue,linkcolor=magenta]{hyperref}

\begin{document}

\title[Quantum correlations in composite systems]{Quantum correlations in composite systems}

\author{J Sperling}
\address{Clarendon Laboratory, University of Oxford, Parks Road, Oxford OX1 3PU, United Kingdom}
\ead{jan.sperling@physics.ox.ac.uk}

\author{E Agudelo}
\address{Arbeitsgruppe Theoretische Quantenoptik, Institut f\"ur Physik, Universit\"at Rostock, D-18051 Rostock, Germany}
\ead{elizabeth.ospina@uni-rostock.de}

\author{I A Walmsley}
\address{Clarendon Laboratory, University of Oxford, Parks Road, Oxford OX1 3PU, United Kingdom}

\author{W Vogel}
\address{Arbeitsgruppe Theoretische Quantenoptik, Institut f\"ur Physik, Universit\"at Rostock, D-18051 Rostock, Germany}

\begin{abstract}
	We study emerging notions of quantum correlations in compound systems.
	Based on different definitions of quantumness in individual subsystems, we investigate how they extend to the joint description of a composite system.
	Especially, we study the bipartite case and the connection of two typically applied and distinctively different concepts of nonclassicality in quantum optics and quantum information.
	Our investigation includes the representation of correlated states in terms of quasiprobability matrices, a comparative study of joint and conditional quantum correlations, and an entanglement characterization.
	It is, for example, shown that our composition approach always includes entanglement as one form of quantum correlations.
	Yet, other forms of quantum correlations can also occur without entanglement.
	Finally, we give an outlook towards multimode systems and temporal correlations.
\end{abstract}

\pacs{
	03.65.-w,	
	03.65.Ud,	
	42.50.-p,	
	03.67.-a	
}

\begin{indented}
	\item[]\today
\end{indented}

\vspace{2pc}
{\it Keywords}: Quantum correlations, nonclassicality, quantum coherence, quantum entanglement

\maketitle
\ioptwocol

\section{Introduction}
	The study of correlations in interacting physical systems is a vital field of research.
	Especially, upcoming quantum technologies urge a redefinition of the notion of correlations to discern classical phenomena from quantum effects.
	This becomes particularly important when quantum systems are joined to form a composite system \cite{KBKMPRS15}.
	For instance, the concept of quantum entanglement, which was described in the seminal work by Einstein, Podolski, and Rosen \cite{EPR35}, emerges from the compound and nonlocal description of at least two quantum particles \cite{HHHH09}.
	Starting from the prominent paradox of Schr\"odinger's cat \cite{S35}, quantum superpositions---in combination with quantum measurements---have been established to be the origin of general quantum correlations which have no counterpart in classical physics.
	Superimposed quantum states explain why quantum systems have the potential to exhibit nonclassical features which cannot be described through classical statistical correlations.
	Consequently, quantum correlations can be applied to implement quantum communication and information protocols \cite{NC10}.

	In order to study nonclassical correlations, a proper concept of a classical state of a quantum system needs to be defined.
	Given the richness of possible physical systems, it is no surprise that various notions of classicality have been established, each of which has a classical analogue in the scenario under study, see, \textit{e.g.}, \cite{P86}.
	For example, coherent states of a quantized radiation field represent the best approximation to a classical electromagnetic wave \cite{S26}.
	Therefore, the Glauber-Sudarshan representation of quantum states of light in terms of such coherent states is the fundamental benchmark to probe nonclassical properties in optical systems \cite{S63,G63,TG65,M86}.
	With respect to this notion, the photon---the quantum particle of an electromagnetic wave \cite{E05}---is clearly a nonclassical state of a light field which can be described as a superposition of coherent states.
	Another example of a nonclassical state stems from the field of quantum information.
	The qubit (quantum bit) introduces the concept of a quantum superposition of classical truth values ``true'' and ``false'' \cite{L04}.
	These superpositions have been shown to lead to a vast number of quantum computation and communication protocols which cannot be implemented using classical information theory \cite{NC10}.
	Examples are quantum key distribution \cite{BB84} or dense coding \cite{BW92}, but also no-go theorems, such as the no-cloning theorem \cite{WZ82}.

	In particular, quantum communication \textit{per se} requires an interplay of at least two subsystems.
	Hence, the concept of nonclassicality has to be generalized to be applicable to compound physical systems.
	The phenomenon of quantum entanglement is one universally accepted type of a nonclassical correlation between multiple degrees of freedom \cite{W89}.
	Especially in highly multipartite systems, the entanglement between individual or ensembles of particles can become rather complex \cite{AFOV08,KL02,HV13,LM13,SSV14,GSVCRTF16} as the relations between the different constituents can exhibit different levels of correlations ranging from uncorrelated over classically correlated to quantum correlated, \textit{i.e.} entangled in the present case.
	The most prominent example for nonequivalent classes of multipartite entanglement is certainly the difference of three qubits in the GHZ or W configuration \cite{DVC00,HSK16}, see also \cite{VSK13} for a generalization.
	However, the detection of entanglement is already challenging when restricting to a bipartite scenario and, consequently, a number of different entanglement tests have been proposed and implemented to uncover such nonclassical correlations, see \cite{HHHH09,GT09} for overviews.

	Another application of quantum correlations is the remote state preparation which is connected to the phenomena of quantum steering \cite{S36,WJD07} and often used when producing (nonclassical) single photons from photon-pair sources via heralding \cite{HM86,SVA14}.
	The underlying conditional realization of quantum states can be a result of entanglement, but it can be also achieved with other nonclassical correlations \cite{OGHW16,BAP16,MZFL16}.
	Therefore, various other concepts of quantum correlations have been exploited \cite{ABC16}.
	For example, the resources needed for the generation of single photons can be completely classified and implemented in terms of nonclassical photon-photon correlations between two separable (\textit{i.e.} disentangled) radiation fields using the Glauber-Sudarshan representation for harmonic oscillators \cite{FP12,ASV13,SBVHBAS15}.

	Despite all those advances in the field of quantum correlations, the urgent question how to treat quantum correlations that arise from distinct notions of nonclassicality of specific physical subsystems---adhered to different concepts of nonclassicality---has not been frequently addressed in the existing literature.

	In this work, we study quantum correlations that emerge in composite quantum systems consisting of parts with diverse notions of nonclassicality.
	In particular, we consider a composed system that consists of a harmonic oscillator and a qubit.
	The notion of nonclassical correlation in the resulting bipartite scenario is defined on the basis of the quantumness of the individual parts.
	Examples of resulting hybrid states highlight the impact of local and global quantum superpositions on the nonclassicality in the bipartite system.
	The implications of joint and conditional quantum correlations are compared.
	Finally, the entanglement of the system under study is discussed and the generalization to multimode systems and to capture temporal correlations is outlined.

	The paper is organized as follows.
	The studied concept of quantum correlations is defined in section \ref{sec:QuantCorr}.
	Different forms of quantum statistical correlations are considered and compared in section \ref{sec:CondJoint}.
	In section \ref{sec:Ent}, we specifically investigate entanglement and we comment on possible extensions of our approach.
	We summarize and conclude in section \ref{sec:Summary}.

\section{Nonclassicality in composed systems}\label{sec:QuantCorr}

	Suppose we have two Hilbert spaces $\mathcal H^{(1)}$ and $\mathcal H^{(2)}$ which describe the two physical subsystems of the compound space $\mathcal H^{(1,2)}=\mathcal H^{(1)}\otimes\mathcal H^{(2)}$.
	Further on, we have two subsets, $\mathcal C^{(j)}\subset\mathcal H^{(j)}$ for $j=1,2$, that consist of pure states which are considered to have classical properties.
	The particular definition of a classical state $|a^{(j)}\rangle\in\mathcal C^{(j)}$ has to be specified and some distinct examples will be given later on.
	A classical pure state in the composite system $\mathcal H^{(1,2)}$ is an uncorrelated state as superpositions would lead to nonclassical/quantum interferences.
	Thus, $\mathcal C^{(1,2)}=\{|a^{(1)}\rangle\otimes|a^{(2)}\rangle:|a^{(1)}\rangle\in\mathcal C^{(1)}\wedge |a^{(2)}\rangle\in\mathcal C^{(2)}\}$ defines the set of classical bipartite states.
	Note that we always assume that states are properly normalized.
	In addition, we indicate the relation to the first and second subsystem as well as the composite system with the superscripts $(1)$, $(2)$, and $(1,2)$, respectively, throughout this work.

	A quantum system that is subjected to an interaction with an environment, or when it is exposed to other imperfections, is not in a pure state.
	Rather, it is a statistical mixture of states.
	Hence, the notion of classical states can be extended to mixed ones.
	That is, any classical mixing of classical pure states yields a (mixed) classical state,
	\begin{eqnarray}
		\hat\rho^{(1,2)}=\int \rmd P_\mathrm{cl.}(a^{(1)},a^{(2)})\, |a^{(1)}\rangle\langle a^{(1)}|\otimes|a^{(2)}\rangle\langle a^{(2)}|,
	\end{eqnarray}
	where $P_\mathrm{cl.}$ is an arbitrary classical probability distribution over the set $\mathcal C^{(1,2)}$.
	Whenever the quantum state of a bipartite system cannot be written in this form, we speak about nonclassical (or quantum) correlations---in contrast to the nonclassicality of the individual subsystems.

	Typically (but not necessarily), nonclassical correlations are related to entanglement.
	Any pure entangled state, decomposed as $\sum_r \psi_r |x^{(1)}_r\rangle \otimes |x^{(2)}_r\rangle$ with linearly independent $|x^{(j)}_r\rangle\in\mathcal H^{(j)}$, is obviously not a product state.
	Therefore, it is not in the set of pure classical states $\mathcal C^{(1,2)}$.
	This holds analogously for mixed entangled states.
	However, we will also discuss nonclassical correlations between the subsystems that are independent of entanglement.

	In the following, we perform a detailed study of the introduced concept for some examples.
	We demonstrate that the quantum superposition principle is the source for nonclassicality within and between the subsystems.
	It is also shown that one can have compound nonclassical states which, however, are classical in their subsystems.

\subsection{A single harmonic oscillator}
	The fundamental notion of nonclassicality in a system of a harmonic oscillator (representing, for example, one optical mode) is given in terms of the Glauber-Sudarshan $P$ representation of a state \cite{S63,G63},
	\begin{eqnarray}\label{eq:glaubersudarshan}
		\hat\rho^{(1)}=\int \rmd^2\alpha\, P(\alpha)\, |\alpha\rangle\langle\alpha|.
	\end{eqnarray}
	In fact, this decomposition is possible for any state via a quasiprobability density $P(\alpha)$.
	The state $\hat\rho^{(1)}$ is classical if and only if $P(\alpha)$ can be interpreted in terms of classical probability theory \cite{TG65,M86}, $P(\alpha)\geq0$.
	If this is not possible, the state refers to as a nonclassical one.
	This means that $P(\alpha)$ is not positive semi-definite in the sense of distributions.
	In general, nonorthogonal coherent states are the classical and pure states in this system \cite{H86},  $\mathcal C^{(1)}=\{|\alpha\rangle:\alpha\in\mathbb C\}$.
	For example, the pure coherent states $|{\pm}\alpha\rangle$ are classical, whereas the superposition $\mathcal N(|{+}\alpha\rangle+|{-}\alpha\rangle)$ ($\mathcal N$ is the normalization constant) is a nonclassical superposition state.

	Independent of the specific quantum characteristics of the state $\hat\rho^{(1)}$, the $P$ function is a real-valued ($P(\alpha)=P(\alpha)^\ast$) and normalized ($\int \rmd^2\alpha\,P(\alpha)=1$) distribution.
	However, the $P$ function is highly singular for many quantum states, such as Fock states and squeezed states \cite{VW06,S16}.
	Consequently, a regularization procedure---known as filtering---has been introduced to verify the nonclassicality via a regular quasiprobability $P_\Omega$ \cite{KV10}, which is given by the convolution
	\begin{eqnarray}\label{eq:filteredP}
		P_\Omega(\alpha)=\int \rmd^2\beta\,P(\beta)\tilde\Omega(\alpha-\beta).
	\end{eqnarray}
	The specific requirements on $\tilde \Omega$---more precisely on its Fourier transform $\Omega$---have been elaborated in \cite{KV10} to guarantee an unambiguous verification of any type of nonclassicality in systems of harmonic oscillators.
	One example of a properly defined filter is the function
	\begin{eqnarray}\label{eq:filter}
		\tilde \Omega(\alpha)=\frac{w^2}{\pi^2}\left[\frac{\sin(w\mathrm{Re}[\alpha])}{w\mathrm{Re}[\alpha]}\right]^2\left[\frac{\sin(w\mathrm{Im}[\alpha])}{w\mathrm{Im}[\alpha]}\right]^2,
	\end{eqnarray}
	which includes the width parameter $0<w\leq \infty$.
	More general classes of filters can be additionally found in \cite{KV14}.
	This regularization approach has been used, for example, to experimentally uncover the negative quasiprobability of a squeezed state \cite{KVHS11,KVCBAP12,ASVKMH15}, which would be impossible with the frequently applied Wigner function.
	Even an experimental quantum-process characterization has been performed based on this filtering technique \cite{RKVGZB13}.

\subsection{A single qubit}
	In contrast to the harmonic oscillator, the classical states of a qubit system are defined through the ability to decompose the quantum state in the diagonal form
	\begin{eqnarray}\label{eq:classicalqubit}
		\hat\rho^{(2)}=P_{0,0}|0\rangle\langle0|+P_{1,1}|1\rangle\langle1|,
	\end{eqnarray}
	where a fixed, orthonormal basis $\{|0\rangle,|1\rangle\}$ is used---eigenvectors of the Pauli matrix $\hat\sigma_z$.
	This means, the state is in a classical statistical mixture of the classical truth values $0$ (``false'') and $1$ (``true''), which are realized with the probabilities $P_{0,0}$ and $P_{1,1}$, respectively.
	This also implies that $\mathcal C^{(2)}=\{|0\rangle,|1\rangle\}$.
	The state $\hat\rho^{(2)}$ is nonclassical if the decomposition \eref{eq:classicalqubit} is impossible, \textit{i.e.} off-diagonal contributions $P_{0,1}|0\rangle\langle 1|+P_{0,1}^\ast|1\rangle\langle 0|$ are required to describe the qubit under study.
	For example, the state $(|0\rangle+|1\rangle)/\sqrt 2$ is nonclassical as it superimposes the classical states for ``true'' and ``false'', which exceeds the limitations of a classical and probabilistic Boolean logic system.

	The quantification of the quantumness in a multi-qubit system has been done in \cite{SV15} as a special example for a general quantification approach using quantum superpositions.
	Moreover, this concept of nonclassicality is typically discussed in the context of quantum coherence for a more general $d$-level system (qudit) \cite{LM14,BCP14}.
	More recently, a quantitative relation between the amount of quantum coherence and entanglement has been found \cite{SSDBA15}.

\subsection{The composite hybrid system}

	In our scenario, the composition $\mathcal C^{(1,2)}=\{|\alpha\rangle\otimes|n\rangle:\alpha\in\mathbb C\wedge n=0,1\}$ defines the set of classical states of the compound bipartite system.
	Since the first system is described in terms of continuous variables and the second one in terms of a discrete variable, such a joint system is called a hybrid system.
	The combined notion of nonclassical correlations allows us to study the origins of different forms of quantum properties.

	Using the previously studied expansions, any hybrid quantum state can be decomposed as \cite{ASCBZV17}
	\begin{eqnarray}\label{eq:hybridstate}
		\hat\rho^{(1,2)}=\int \rmd^2\alpha\sum_{n,n'} P_{n,n'}(\alpha)\, |\alpha\rangle\langle\alpha|\otimes|n\rangle\langle n'|,
	\end{eqnarray}
	where $P_{n,n'}(\alpha)=P_{n',n}(\alpha)^\ast$.
	Therefore, the combined notion of a mixed classical state implies that $P_{0,1}(\alpha)=0$, $P_{0,0}(\alpha)\geq 0$, and $P_{1,1}(\alpha)\geq 0$ has to hold for any classical hybrid state.
	For convenience, we can define a $P$ matrix, $\boldsymbol P(\alpha)=(P_{n,n'}(\alpha))_{n,n'=0,1}$, to describe the compound system and for generalizing the Glauber-Sudarshan $P$ distribution.
	This $P$ matrix can be filtered as shown for the single-mode case in equation \eref{eq:filteredP}.
	This concept was introduced in \cite{ASCBZV17}.

	Before exploiting this joint description in more detail, it is important to stress a major difference in the notion of classical qubits \eref{eq:classicalqubit} and the case of the Glauber-Sudarshan representation \eref{eq:glaubersudarshan}.
	Because of $P_{0,1}\neq0$ for nonclassical qubit states, the decomposition \eref{eq:classicalqubit} is not possible for them even if one would allow for negative or complex quasiprobabilities $P_{n,n}$ ($n=0,1$).
	In contrast the expansion in \eref{eq:glaubersudarshan} applies to all quantum states of an harmonic oscillator.
	In general, any system with pure, classical reference states could include the following types of nonclassical states:
	\begin{itemize}
		\item[(i)] States that can be represented in terms of pseudo-mixtures of pure states (\textit{e.g.} nonclassical states of the harmonic oscillator) and
		\item[(ii)] states that cannot be represented in terms of such a diagonal-representation of classical states (\textit{e.g.} nonclassical qubit states).
	\end{itemize}
	This means that for nonclassical states of the form (i), a quasiprobability representation in terms of classical states exists which is impossible for quantum states of the type (ii).

	Let us consider some first elementary examples of bipartite nonclassical states to highlight different forms of quantumness,
	\numparts
	\begin{eqnarray}
		|\phi\rangle=&\frac{|\alpha\rangle+|{-}\alpha\rangle}{\sqrt{2(1+\rme^{-2|\alpha|^2})}}\otimes|0\rangle,
		\\
		|\psi\rangle=&|\alpha\rangle\otimes\frac{|0\rangle+|1\rangle}{\sqrt{2}},
		\\\label{eq:pureexample3}
		|\chi\rangle=&\frac{|\alpha\rangle\otimes |0\rangle+|{-}\alpha\rangle\otimes |1\rangle}{\sqrt 2},
	\end{eqnarray}
	\endnumparts
	for $\alpha\neq0$.
	The state $|\phi\rangle$ is solely nonclassical with respect to the first subsystem and it corresponds to the case of nonclassicality (i).
	That is, its density operator $|\phi\rangle\langle\phi|$ is described with a $P$ matrix whose elements are $P_{1,0}(\alpha)=0$ and $P_{0,0}(\alpha)\ngeq0$ (as well as $P_{1,1}(\alpha)=0$).
	Hence, $|\phi\rangle$ has no classical harmonic oscillator analogue.
	An example for the type of nonclassicality of the case (ii) is the state $|\psi\rangle$.
	It only shows nonclassicality in the second, qubit subsystem and has a non-vanishing off-diagonal contribution in the $P$ matrix, $P_{0,1}(\alpha)=\delta(\alpha)/2\neq0$ (note, $P_{0,0}(\alpha)=P_{1,1}(\alpha)=P_{0,1}(\alpha)\geq0$).
	So far, we solely considered the local superposition states.

	Finally, the state $|\chi\rangle$ in equation \eref{eq:pureexample3} is entangled and, therefore, shows nonclassical correlations between the two components of the hybrid system.
	This is of some importance, because its single-mode reduced states are classical mixtures of pure classical states in their respective subsystems.
	That is, we have for the second subsystem $\mathrm{tr}_1|\chi\rangle\langle\chi|=(|0\rangle\langle0|+|1\rangle\langle1|)/2$, in the limit $\langle \alpha|{-}\alpha\rangle=\rme^{-2|\alpha|^2}\approx0$ for $|\alpha|\gg 1$, and $\mathrm{tr}_2|\chi\rangle\langle\chi|=(|\alpha\rangle\langle\alpha|+|{-}\alpha\rangle\langle{-}\alpha|)/2$ for the first subsystem.
	The entanglement of $|\chi\rangle$---being the present form of nonclassical correlations---is a result of the global superposition of the two classical hybrid states $|\alpha\rangle\otimes|0\rangle$ and $|{-}\alpha\rangle\otimes|1\rangle$.
	The role of the superpositions for entanglement has been also experimentally determined in \cite{GPMMSVP14}.
	Let us also mention that combinations of the quantum superposition principle on the local and global level can be also used to get intermediate levels of nonclassicality.
	Those combine nonclassicality in the individual subsystems and nonclassical correlations, for example, $\mathcal N([|\alpha\rangle+|{-}\alpha\rangle]\otimes[|0\rangle+|1\rangle]+[|\alpha\rangle+\rmi|{-}\alpha\rangle]\otimes[|0\rangle+\rmi|1\rangle])$.

\subsection{Example: Mixed cat states}
	As a generalization of a previous example, we study a family Schr\"odinger cat states \cite{S35},
	\begin{eqnarray}\label{eq:catstate}
		|\chi_\varphi\rangle=&\frac{|\alpha_0\rangle\otimes |0\rangle+\rme^{\rmi\varphi}|{-}\alpha_0\rangle\otimes |1\rangle}{\sqrt 2},
	\end{eqnarray}
	which is particularly parametrized through the phase $0\leq\varphi<2\pi$ between the classical terms.
	Here, we restrict ourselves to real-valued and non-negative coherent amplitudes, $\alpha_0=\alpha_0^\ast\geq0$.
	It is also worth mentioning that relations between phase-dependent properties and entanglement of similar states have been studied in the presence of imperfections, such as fluctuating losses, in \cite{BSV15}.

	Let us assume that the phase can be subjected to a randomization according to a $2\pi$-periodic Gaussian distribution with a variance $\sigma^2$.
	This results in the mixed state
	\begin{eqnarray}\label{eq:mixedcatstate}
		\hat\varrho_\sigma&=&\int_{0}^{2\pi}\rmd\varphi\,\sum_{k\in\mathbb Z}\frac{\exp\left(-\frac{(\varphi-2\pi k)^2}{2\sigma^2}\right)}{\sqrt{2\pi\sigma^2}}|\chi_\varphi\rangle\langle\chi_\varphi|
		\\\nonumber
		&=&\frac{1}{2}\Big(|\alpha_0\rangle\langle\alpha_0|\otimes|0\rangle\langle 0|
		+|{-}\alpha_0\rangle\langle{-}\alpha_0|\otimes|1\rangle\langle 1|\Big)
		\\\nonumber
		&&{+}\frac{\tau_\sigma}{2}\Big(|{-}\alpha_0\rangle\langle\alpha_0|\otimes|1\rangle\langle 0|
		+|\alpha_0\rangle\langle{-}\alpha_0|\otimes|0\rangle\langle 1|\Big),
	\end{eqnarray}
	with the real-valued and non-negative parameter
	\begin{eqnarray}
		\tau_\sigma=&\int_{0}^{2\pi}\rmd\varphi\,\sum_{k\in\mathbb Z}\frac{\exp\left(-\frac{(\varphi-2\pi k)^2}{2\sigma^2}{+}\rmi\varphi\right)}{\sqrt{2\pi\sigma^2}}=\rme^{-\sigma^2/2}.
	\end{eqnarray}
	Note that non-Gaussian phase randomizations could be considered in a similar manner.
	As we will show in this work, the quantum properties of this state are determined through the parameter $\tau_\sigma$.
	Especially from the bounds $\tau_0=1$ and $\tau_\infty=0$, we can already observe that the former yields the pure superposition state \eref{eq:catstate}, for $\varphi=0$, and the latter results in the classical mixture
	\begin{eqnarray}
		\hat\rho_\infty=\frac{1}{2}\Big(|\alpha_0\rangle\langle\alpha_0|{\otimes}|0\rangle\langle 0|+|{-}\alpha_0\rangle\langle{-}\alpha_0|{\otimes}|1\rangle\langle 1|\Big).
	\end{eqnarray}

\begin{figure}[ht]
	\centering
	\includegraphics[width=\linewidth]{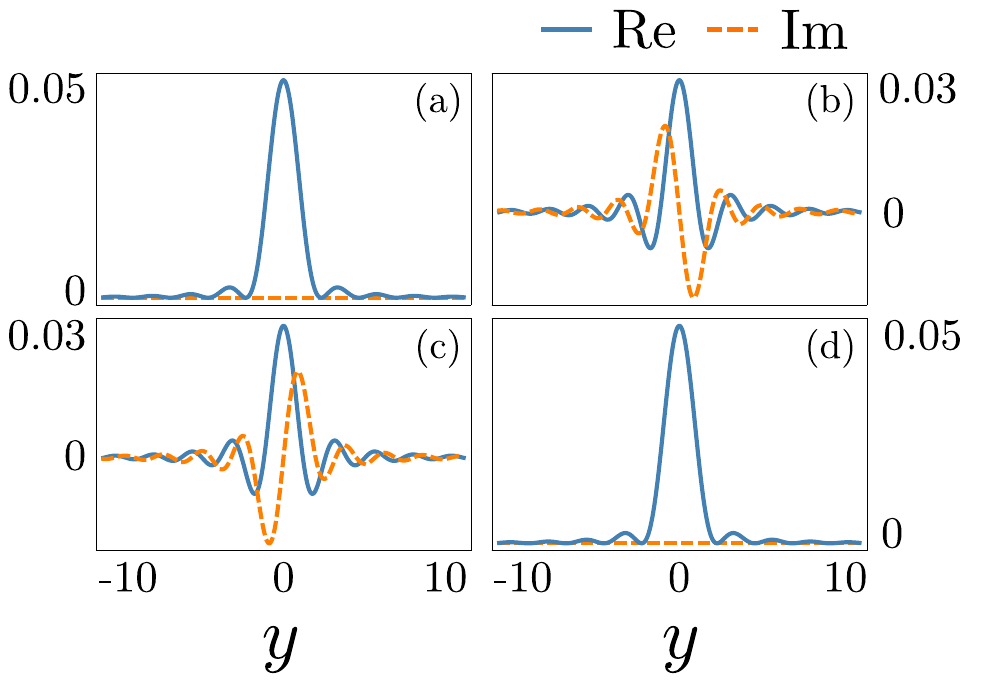}
	\caption{
		Real (solid, blue) and imaginary (dashed, orange) part of the cross-sections ($x=\mathrm{Re}[\alpha]=0$), of the regularized $P$ matrix representation of the hybrid state $\hat\varrho_\sigma$ as a function of $y=\mathrm{Im}[\alpha]$.
		We used the filter width $w=1.5$, a Gaussian dephasing of $\sigma=0.5$, and a coherent amplitude $\alpha_0=1$.
		The panels (a) and (d) show the diagonal elements \eref{eq:regPexample0} and \eref{eq:regPexample1}, respectively.
		The panels (b) and (c) show the off-diagonal elements \eref{eq:regPexampleOff} and its complex conjugate, respectively.
	}\label{Fig:Pmatrix}
\end{figure}

	Let us give the $P$ matrix representation for the state \eref{eq:mixedcatstate}.
	For this reason, we recall that a coherent state can be written in the form $|\alpha\rangle=\exp(-|\alpha|^2/2+\alpha\,\hat a^\dag)|0\rangle$ \cite{VW06}, where $\hat a^\dag$ is the bosonic creation operator.
	Using additionally $\alpha=x+\rmi y$, we can write
	\begin{eqnarray*}
		&\int \rmd^2\alpha\,\delta(\mathrm{Re}\,\alpha)\delta(\mathrm{Im}\,\alpha+\rmi\alpha_0)|\alpha\rangle\langle\alpha|
		\\{=}&\int \rmd x\,\rmd y\,\delta(x)\delta(y{+}\rmi\alpha_0)\rme^{{-}x^2{-}y^2}\rme^{(x{+}\rmi y)\hat a^\dag}|0\rangle\langle0|\rme^{(x{-}\rmi y)\hat a}
		\\=&\rme^{\alpha_0^2}\rme^{\alpha_0\hat a^\dag}|0\rangle\langle0|\rme^{-\alpha_0\hat a}=\rme^{2\alpha_0^2}|\alpha_0\rangle\langle{-}\alpha_0|.
	\end{eqnarray*}
	Thus, the state $\hat\varrho_\sigma$ has the $P$ matrix representation
	\begin{eqnarray}
		&&\boldsymbol P(\alpha)=
		\\\nonumber&&
		\left(\begin{array}{cc}
			\frac{1}{2}\delta(\mathrm{Re}[\alpha]{-}\alpha_0)\delta(\mathrm{Im}[\alpha]) & P_{0,1}(\alpha)^\ast\\
			P_{0,1}(\alpha) & \frac{1}{2}\delta(\mathrm{Re}[\alpha]{+}\alpha_0)\delta(\mathrm{Im}[\alpha])
		\end{array}\right)\!,
	\end{eqnarray}
	with the off-diagonal element
	\begin{eqnarray}
		P_{0,1}(\alpha)=\tau_\sigma\frac{\rme^{-2\alpha_0^2}}{2}\delta(\mathrm{Re}\,\alpha)\delta(\mathrm{Im}\,\alpha+\rmi\alpha_0).
	\end{eqnarray}
	Finally, the convolution with the filter \eref{eq:filter} yields the regularized matrix elements
	\numparts
	\begin{eqnarray}
		\label{eq:regPexample0}
		P_{\Omega;0,0}(\alpha)=&\frac{1}{2}\frac{w^2}{\pi^2}\left[\frac{\sin(w[x-\alpha_0])}{w[x-\alpha_0]}\right]^2\left[\frac{\sin(wy)}{wy}\right]^2,
		\\\label{eq:regPexample1}
		P_{\Omega;1,1}(\alpha)=&\frac{1}{2}\frac{w^2}{\pi^2}\left[\frac{\sin(w[x+\alpha_0])}{w[x+\alpha_0]}\right]^2\left[\frac{\sin(wy)}{wy}\right]^2,
		\\\label{eq:regPexampleOff}
		P_{\Omega;0,1}(\alpha)=&\tau_\sigma\frac{\rme^{-2\alpha_0^2}}{2}\frac{w^2}{\pi^2}\left[\frac{\sin(wx)}{wx}\right]^2
		\nonumber\\&\times
		\left[\frac{\sin(w[y+\rmi\alpha_0])}{w[y+\rmi\alpha_0]}\right]^2,
	\end{eqnarray}
	\endnumparts
	where we used $x=\mathrm{Re}(\alpha)$ and $y=\mathrm{Im}(\alpha)$.
	Let us emphasize that the off-diagonal component $P_{\Omega;0,1}(\alpha)$ is a complex-valued function.
	The elements of the filtered $P$ matrix for the state $\hat\varrho_\sigma$ are shown in Fig. \ref{Fig:Pmatrix} for $x=0$.
	We clearly see that the off-diagonal elements are not zero and the diagonal ones are non-negative.

\section{Joint and conditional quantum correlations}\label{sec:CondJoint}

\begin{figure}[ht]
	\centering
	\includegraphics[width=0.75\linewidth]{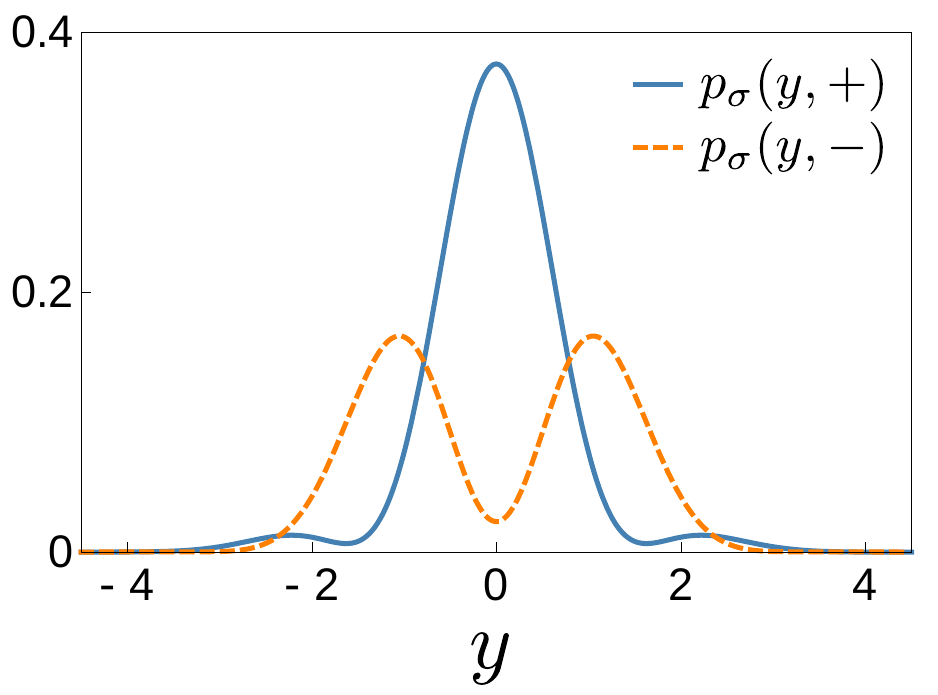}
	\caption{
		Joint probability distribution \eref{eq:jointdistMixedCat} for the hybrid state $\hat\varrho_\sigma$ for $\sigma= 0.5$ and $\alpha_0=1$.
		The behavior of the joint probability distribution $p(y,\pm)$ is shown for the outcomes ``$+$'' (blue, solid) and ``$-$'' (orange, dashed) as a function of the momentum values $y$.
	}\label{Fig:pmdist}
\end{figure}

	In order to access the nonclassicality of the subsystems or the nonclassical correlations between them, measurements have to be performed.
	Let us explore the measurement of the momentum $\hat y=(\hat a-\hat a^\dag)/\rmi$ in the first mode and the measurement of the Pauli matrix $\hat\sigma_x=|0\rangle\langle 1|+|1\rangle\langle 0|$ in the second degree of freedom.
	The detection process can be characterized through its eigenvectors, $\hat y|y\rangle=y|y\rangle$ (eigenvalue $y\in\mathbb R$) and $\hat\sigma_x|\pm\rangle=\pm|\pm\rangle$.
	We have the following projections:
	\begin{eqnarray}
		\langle\pm|0\rangle=\frac{1}{\sqrt 2},\quad
		\langle\pm|1\rangle=\pm\frac{1}{\sqrt 2},
	\end{eqnarray}
	and \cite{VW06}
	\begin{eqnarray}
		\langle y|\alpha\rangle=\frac{\exp\left(-\frac{1}{4}y^2-\rmi\alpha y-\frac{|\alpha|^2-\alpha^2}{2}\right)}{\sqrt[4]{2\pi}}.
	\end{eqnarray}
	For instance for pure classical states $|\alpha\rangle\otimes|n\rangle\in\mathcal C^{(1,2)}$ (with $\alpha\in\mathbb C$ and $n=0,1$), we expect a product probability distribution of the form
	\begin{eqnarray}
		p(y,\pm)=|\langle y|\alpha\rangle|^2\cdot|\langle \pm|n\rangle|^2=\frac{\rme^{-[y-2\mathrm{Im}(\alpha)]^2/2}}{\sqrt{2\pi}}\cdot\frac{1}{2},
	\end{eqnarray}
	for the possible measurement outcomes.
	Mixed classical states will result in the corresponding mixtures of the statistics for pure, classical states.
	For our hybrid state \eref{eq:mixedcatstate}, the joint probability distribution reads
	\begin{eqnarray}\label{eq:jointdistMixedCat}
		p_\sigma(y,\pm)=\frac{\rme^{-y^2/2}}{2\sqrt{2\pi}}\left(1\pm\tau_\sigma\cos[2\alpha_0y]\right).
	\end{eqnarray}
	This distribution is depicted in Fig. \ref{Fig:pmdist} for different outcomes of the Pauli matrix measurements and the momentum measurement.
	
	The considered measurements consists of only two local observables $\hat y\otimes\hat 1$ and $\hat 1\otimes \hat\sigma_x$.
	Therefore, no entanglement can be detected with these measurements alone because Schwarz' theorem applies, \textit{i.e.} $[\hat y\otimes\hat 1,\hat 1\otimes \hat\sigma_x]=0$.
	Thus, any form of quantumness that will be identified in this section is not a result of entanglement but presents other kinds of nonclassical correlation.

	From the jointly measured statistics $p(y,\pm)$ of an arbitrary state, one can compute the marginal probabilities, $p(y)=p(y,+)+p(y,-)$ and $p(\pm)=\int \rmd y\, p(y,\pm)$, as well as the conditional probabilities,
	\begin{eqnarray}
		p(y|\pm)=\frac{p(y,\pm)}{p(\pm)}
		\quad \mathrm{and} \quad
		p(\pm|y)=\frac{p(y,\pm)}{p(y)}.
	\end{eqnarray}
	As the joint probability distribution is well-known and frequently studied, let us make some some remarks on the conditional distributions.
	The conditional statistics can be associated with conditional states, which are in our case
	\numparts
	\begin{eqnarray}
		\label{eq:conditionedhybridstate1}
		&\hat\varrho_\sigma|_{y}=\frac{\mathrm{tr}_1[\hat \rho_\sigma(|y\rangle\langle y|\otimes\hat 1)]}{\mathrm{tr}_{1,2}[\hat \rho_\sigma(|y\rangle\langle y|\otimes\hat 1)]}
		\\\nonumber
		=&\frac{|0\rangle\langle0|+|1\rangle\langle1|+\tau_\sigma \rme^{2\rmi\alpha_0y}|1\rangle\langle 0|+\tau_\sigma \rme^{-2\rmi\alpha_0y}|0\rangle\langle 1|}{2},
	\end{eqnarray}
	and
	\begin{eqnarray}
		\label{eq:conditionedhybridstate2}
		&\hat\varrho_\sigma|_{\pm}=\frac{\mathrm{tr}_2[\hat \rho_\sigma(\hat 1\otimes|\pm\rangle\langle \pm|)]}{\mathrm{tr}_{1,2}[\hat \rho_\sigma(\hat 1\otimes|\pm\rangle\langle \pm|)]}
		\\\nonumber
		=&\frac{|\alpha_0\rangle\langle\alpha_0|{+}|{-}\alpha_0\rangle\langle{-}\alpha_0|\pm\tau_\sigma(|{-}\alpha_0\rangle\langle\alpha_0|{+}|\alpha_0\rangle\langle{-}\alpha_0|)}{2(1\pm\tau_\sigma \rme^{-2\alpha_0^2})}.
	\end{eqnarray}
	\endnumparts
	These conditional states give the conditional probabilities $p_\sigma(\pm|y)$ and $p_\sigma(y|\pm)$ when measuring $\hat\sigma_x$ and $\hat y$, respectively.
	It is also worth mentioning that pure classical states are product states, \textit{i.e.} $p(y,\pm)=p(y)p(\pm)$.
	Thus, their conditional statistics takes the forms $p(y|\pm)=p(y)$ and $p(\pm|y)=p(\pm)$.
	In the remainder of this section, we study nonclassical correlations based on different probability distributions, \textit{i.e.} joint and conditional statistics.

\subsection{Cross-correlations}
	For studying joint correlations, let us consider the following matrix of second-order moments:
	\begin{eqnarray}\label{eq:MDef}
		M=\left(\begin{array}{ccc}
			\langle \hat 1\otimes \hat 1\rangle & \langle \hat y\otimes \hat 1\rangle & \langle \hat 1\otimes \hat \sigma_x\rangle
			\\
			\langle \hat y\otimes \hat 1\rangle & \langle \hat y^2\otimes \hat 1\rangle & \langle \hat y\otimes \hat \sigma_x\rangle
			\\
			\langle \hat 1\otimes \hat \sigma_x\rangle & \langle \hat y\otimes \hat \sigma_x\rangle & \langle \hat 1\otimes \hat \sigma_x^2\rangle
		\end{array}\right).
	\end{eqnarray}
	Note that $\langle \hat 1\otimes \hat 1\rangle=1$ and $\hat\sigma_x^2=\hat 1$.
	From this matrix, we can determine the following nontrivial principal minors:
	\numparts
	\begin{eqnarray}
		\mu^{(1)}&=&\langle (\Delta\hat y)^2\otimes\hat 1\rangle,
		\\
		\mu^{(2)}&=&\langle \hat 1\otimes (\Delta\hat \sigma_x)^2\rangle,
		\\\nonumber
		\mu^{(1,2)}&=&\langle (\Delta\hat y)^2\otimes\hat 1\rangle\langle \hat 1\otimes (\Delta\hat \sigma_x)^2\rangle
		\\
		&&-\langle (\Delta\hat y)\otimes (\Delta\hat \sigma_x)\rangle^2.
	\end{eqnarray}
	\endnumparts
	The $2\times2$ sub-determinants $\mu^{(j)}$ ($j=1,2$) are the variances for the individual subsystems and $\mu^{(1,2)}=\det(M)$ is the cross-correlation of the joint statistics.

	Determinants satisfy the mixing property $\det[qA+(1-q)B]\geq \min\{\det(A),\det(B)\}$ for all $0\leq q\leq 1$ and positive semidefinite matrices $A$ and $B$.
	Further on, the matrix \eref{eq:MDef} is positive semidefinite as it can be written as $M=\langle \hat v\hat v^\dag\rangle$, where $\hat v=(\hat 1\otimes\hat 1,\hat y\otimes\hat 1,\hat 1\otimes\hat \sigma_x)^{\rm T}$.
	Therefore, it is sufficient to analyze the bounds of the above minors for pure classical states.
	If those bounds are violated, we have certified nonclassicality.
	For the state $|\alpha\rangle\otimes|n\rangle\in\mathcal C^{(1,2)}$, we directly compute $\mu^{(1)}=\mu^{(2)}=\mu^{(1,2)}=1$.
	This means, for arbitrary classical states holds $\mu^{(1)}\geq 1$ and a violation $\mu^{(1)}<1$ refers to as squeezing.
	Also, the classical states implies $\mu^{(2)}\geq 1$; note that the maximal variance of $\hat \sigma_x$ is one and $\mu^{(2)}\geq1$ can be relaxed to $\mu^{(2)}=1$ for classical states.
	Finally, $\mu^{(1,2)}\geq 1$ is the bound for classical cross-correlations.

	We find for our hybrid state $\hat\varrho_\sigma$ in equation \eref{eq:mixedcatstate} the following principal minors:
	\begin{equation}
		\mu_\sigma^{(1)}=1
		\quad\mathrm{and}\quad
		\mu_\sigma^{(2)}=1-\tau_\sigma^2\rme^{-4\alpha_0^2}=\mu_\sigma^{(1,2)}.
	\end{equation}
	Hence, we cannot observe squeezing $\mu_\sigma^{(1)}\geq 1$, but we have a nonclassical qubit subsystem $\mu_\sigma^{(2)}<1$ (for $\tau_\sigma^2\neq0$).
	We also have nonclassical cross-correlations from the joint statistics, $\mu_\sigma^{(1,2)}<1$.
	However, this correlation is fully determined through the nonclassicality of the qubit since $\mu_\sigma^{(1,2)}=\mu_\sigma^{(2)}$ for the observables under study.

\subsection{Conditional variances}
	After considering the joint statistics, let us focus on the conditional statistics.
	As mentioned earlier, conditional quantum correlations---despite their importance---are typically not explicitly studied.
	Conditional states and statistics are considered to generate single-mode nonclassical states from bipartite quantum correlated states \cite{SVA14}.
	Such a method can be used to generate squeezed states \cite{VVS06,MZFHTYMF17}.
	The recent experimental application \cite{ASCBZV17} even directly relates to the quantum correlated state \eref{eq:pureexample3}.
	Other applications of conditional correlations are connected to the mutual information \cite{YHW08}, conditional uncertainties \cite{SAPA17}, and quantum randomness \cite{K16}.
	We also compared the quantum correlations and directly verified the either joint or conditional nonclassicality for different experimentally realized states \cite{SBDBJDVW16}.

\begin{figure}[ht]
	\centering
	\includegraphics[width=0.75\linewidth]{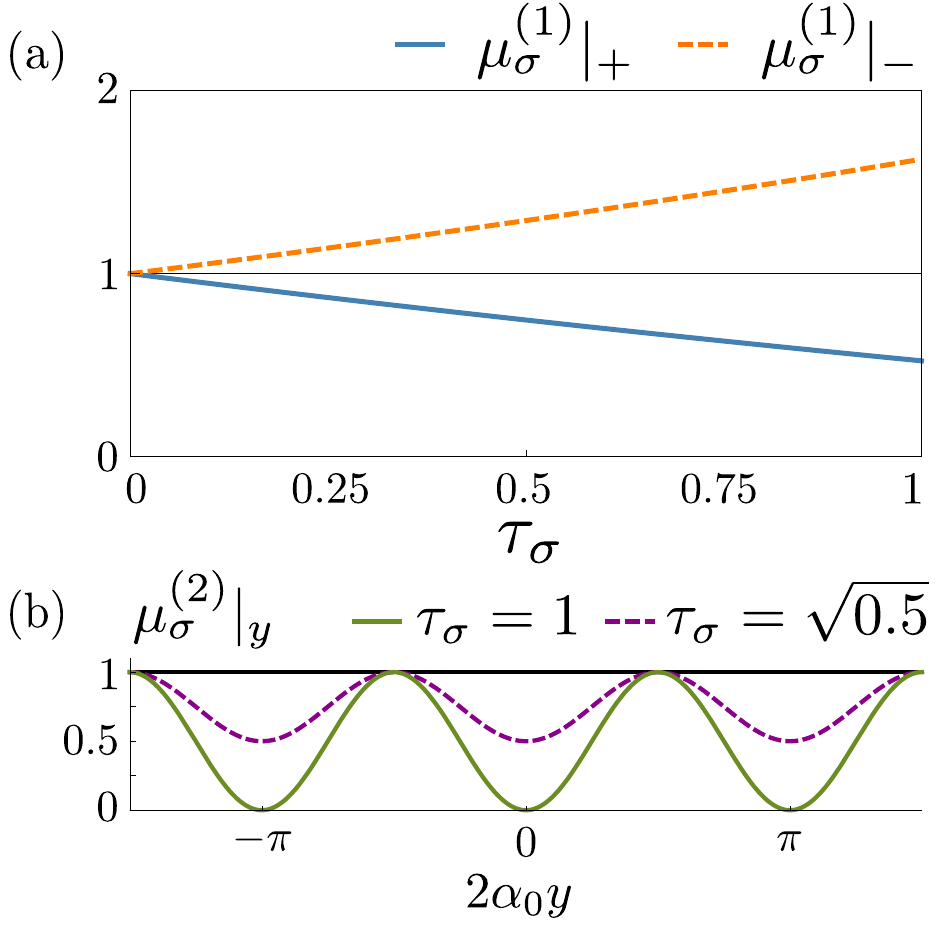}
	\caption{
		Conditional moments for the hybrid state $\hat\varrho_\sigma$ for $\alpha_0=1$ are plotted.
		In the top panel (a), we show the conditional variances  as a function of the parameter $\tau_\sigma$ for the momentum $\hat y$ depending on the outcomes of the $\hat\sigma_x$ measurement (solid/blue and dashed/orange for the $+$ and $-$  case, respectively).
		Conditional squeezing can be observed for the conditioning to $+$.
		In the bottom panel (b), we study the conditional variance of the Pauli matrix depending on the outcome $y$ of the momentum measurement.
		The values $\mu^{(2)}_\sigma|_y<1$ identify nonclassicality of the conditional qubit state for different dephasing levels $\sigma$.
	}\label{Fig:moments}
\end{figure}

	As done before, we can find the following bounds for classical states:
	For the variance of $\hat y$ conditioned to $\pm$ for the $\hat\sigma_x$ measurement holds $\mu^{(1)}|_{\pm}\geq 1$ and analogously holds $\mu^{(2)}|_{y}=1$.
	For our conditioned hybrid state in equations \eref{eq:conditionedhybridstate1} and \eref{eq:conditionedhybridstate2}, we get the following conditional variances:
	\numparts
	\begin{eqnarray}
		\mu^{(1)}_\sigma|_\pm=1\mp\frac{4\alpha_0^2\tau_\sigma \rme^{-2\alpha_0^2}}{1\pm\tau_\sigma \rme^{-2\alpha_0^2}}
	\end{eqnarray}
	and
	\begin{eqnarray}
		\mu^{(2)}_\sigma|_y=1-\tau_\sigma^2\cos^2(2\alpha_0y).
	\end{eqnarray}
	\endnumparts
	Those conditional moments are depicted in Fig. \ref{Fig:moments}.
	Conditioned to the measurement outcome $+$, we find that we can observe squeezing as $\mu^{(1)}_\sigma|_+<1$.
	Let us emphasize that this squeezing was not detected with the joint statistics in the previous subsection.
	Another difference to the joint variances is that we cannot detect the qubit nonclassicality for certain conditionings of the momentum measurement, $\mu^{(2)}_\sigma|_y=1$ for cases with $\cos(2\alpha_0y)=0$.
	
\section{Entanglement and beyond}\label{sec:Ent}

\subsection{Entanglement}
	In this subsection, we are going to study the nonclassical correlation of entanglement.
	Again, we will focus on our hybrid system consisting of one harmonic oscillator and one qubit and investigate the resulting hybrid entanglement.
	Let us point out that entanglement is always a nonclassical effect in our construction of compound systems.
	Even if the individual sets of classical states $\mathcal C^{(j)}$ include all single-mode states, the set $\mathcal C^{(1,2)}$ of jointly classical states solely includes product states.

	A number of interesting relations between entanglement and other notions of nonclassicality have been studied.
	For example, the emission of nonclassical light from a localized source in many directions automatically implies multipartite entanglement \cite{GV14}.
	Also, a certain amount of single-mode nonclassicality of a harmonic oscillator yields---after the light is split on a beam splitter---the same amount of entanglement \cite{VS14}.
	A similar connection has been found for the notion of quantum coherence \cite{SSDBA15} that relates to our qubit system.
	Remarkably, a universal conversion approach of local nonclassicality to entanglement has been formulated in \cite{KSP16}.

	For the detection of entanglement, we introduced the so-called separability eigenvalue equations \cite{SV09,SV13}, which read for a bipartite observable $\hat L$ as
	\numparts\label{eq:sepevaleqs}
	\begin{eqnarray}\label{eq:sepevaleqs1}
		\hat L_{a^{(2)}}|a^{(1)}\rangle=g|a^{(1)}\rangle,
		\\\label{eq:sepevaleqs2}
		\hat L_{a^{(1)}}|a^{(2)}\rangle=g|a^{(2)}\rangle,
	\end{eqnarray}
	\endnumparts
	where $\langle a^{(j)}|a^{(j)}\rangle=1$ and $\hat L_{a^{(1)}}=\mathrm{tr}_1[\hat L(|a^{(1)}\rangle\langle a^{(1)}|\otimes\hat 1)]$ (analogously $\hat L_{a^{(2)}}$).
	The minimal and maximal separability eigenvalues $g$ determine the bounds for the expectation value of $\hat L$ for separable states,
	\begin{eqnarray}\label{eq:entanglementcond}
		\min\{g\}\leq\langle\hat L\rangle_{\rm sep.}\leq\max\{g\}.
	\end{eqnarray}
	Whenever these bounds are violated, the state is entangled.
	Also note that we have shown that for any entangled state exists such an observable $\hat L$ which can detect its entanglement.
	This approach of separability eigenvalue equations to construct entanglement tests has been used to study the emission of entangled light from semiconductor structures \cite{PFSV12,PFSV13}, to formulate entanglement quasiprobabilities \cite{SV09a,SV12} (see also \cite{STV98} in this context), or to experimentally characterize highly multimode entanglement beyond the capabilities of earlier approaches \cite{GSVCRTF15,GSVCRTF16}.

	To detect entanglement in the system under study, let us consider the observable
	\begin{eqnarray}
		\hat L=|\alpha_0\rangle\langle{-}\alpha_0|\otimes|0\rangle\langle1|+|{-}\alpha_0\rangle\langle\alpha_0|\otimes|1\rangle\langle0|,
	\end{eqnarray}
	which addresses the interference terms between the two modes.
	The solution of the second equation \eref{eq:sepevaleqs2} can be directly formulated,
	\begin{eqnarray}\label{eq:sepsolution2}
		|a^{(2)}\rangle=\frac{1}{\sqrt 2}|0\rangle\pm\frac{1}{\sqrt 2}
			\frac{\langle a^{(1)}|{-}\alpha_0\rangle}{|\langle a^{(1)}|{-}\alpha_0\rangle|}
			\frac{\langle \alpha_0|a^{(1)}\rangle}{|\langle \alpha_0|a^{(1)}\rangle|}
		|1\rangle.
	\end{eqnarray}
	With this, the first equation \eref{eq:sepevaleqs1} reads
	\begin{eqnarray*}
		\hat L_{a^{(2)}}|a^{(1)}\rangle&{=}&{\pm}\frac{1}{2}
		\frac{\langle \alpha_0|a^{(1)}\rangle}{|\langle \alpha_0|a^{(1)}\rangle|}
		\frac{\langle {-}\alpha_0|a^{(1)}\rangle}{|\langle {-}\alpha_0|a^{(1)}\rangle|}
		\\&&{\times}\Big[
			\langle a^{(1)}|{-}\alpha_0\rangle|\alpha_0\rangle
			+\langle a^{(1)}|\alpha_0\rangle|{-}\alpha_0\rangle
		\Big]
		\\&{=}& g|a^{(1)}\rangle.\nonumber
	\end{eqnarray*}
	Using a proper rescaling allows us to get the solution from this equation by solving the simplified problem
	$\langle a^{(1)}|{-}\alpha_0\rangle|\alpha_0\rangle+\langle a^{(1)}|\alpha_0\rangle|{-}\alpha_0\rangle=\gamma|a^{(1)}\rangle$ together with the ansatz $|a^{(1)}\rangle=\mathcal N[|\alpha_0\rangle+s|{-}\alpha_0\rangle]$.
	Equating coefficients and inserting the resulting two equations into each other to eliminate $\gamma$ yields $\mathrm{Im}(s)=0$ and $|s|^2=1$.
	Thus, we get
	\begin{eqnarray}\label{eq:sepsolution1}
		|a^{(1)}\rangle=\mathcal N(|\alpha_0\rangle\pm'|{-}\alpha_0\rangle),
	\end{eqnarray}
	where $\pm'$ means that this sign is independent from the one in equation \eref{eq:sepsolution2}.
	This also determines the parameters for the solution of $|a^{(2)}\rangle$ in equation \eref{eq:sepsolution2} and the separability eigenvalues
	\begin{eqnarray}
		g=\pm\frac{1\pm'\rme^{-2\alpha_0^2}}{2}.
	\end{eqnarray}

\begin{figure}[ht]
	\centering
	\includegraphics[width=0.75\linewidth]{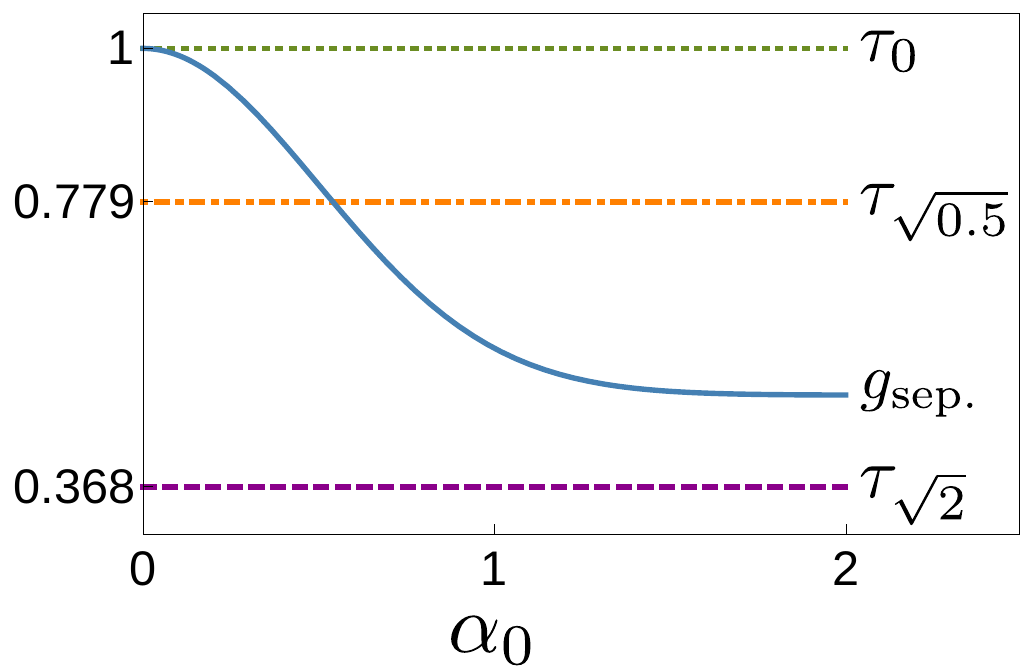}
	\caption{
		The entanglement condition \eref{eq:finalenttest} applied to the hybrid states $\hat\varrho_0$, $\hat\varrho_{\sqrt{0.5}}$, and $\hat\varrho_{\sqrt{2}}$ is shown as a function of the coherent amplitude $\alpha_0$.
		The maximal expectation value for separable state (solid, blue line labeled as $g$) is violated for no dephasing (dotted, green curve) and for $\alpha_0>0.54$ in case of the moderate dephasing (dot-dashed, orange curve).
		In case of higher phase noise (dashed, violet curve), no entanglement can be detected with the observable $\hat L$ under study.
	}\label{Fig:gtau}
\end{figure}

	From the maximum and minimum of the separability eigenvalue equations, we find with \eref{eq:entanglementcond} that for all separable states holds
	\begin{eqnarray}\label{eq:finalenttest}
		\left|\langle\hat L\rangle_{\rm sep.}\right|\leq\frac{1+\rme^{-2\alpha_0^2}}{2}=g_{\rm sep.}.
	\end{eqnarray}
	For our considered state in equation \eref{eq:mixedcatstate}, we get the expectation value $\langle\hat L\rangle=\tau_\sigma=\rme^{-\sigma^2/2}$.
	This means that we can detect entanglement for any
	\begin{eqnarray}
		\tau_\sigma>\frac{1+\rme^{-2\alpha_0^2}}{2}.
	\end{eqnarray}
	In Fig. \ref{Fig:gtau}, the resulting detection of entanglement is shown for our hybrid system.
	
\subsection{Multimode generalizations}
	A generalization of our approach to multimode systems is straightforward.
	For the $N$-partite scenario, we consider the sets $\mathcal C^{(j)}$ of pure classical states for $j=1,\ldots,N$.
	The compound set of classical states is consequently defined as
	\begin{eqnarray}
		\nonumber
		\mathcal C^{(1,\ldots,N)}&=&\{|a^{(1)}\rangle\otimes\cdots\otimes|a^{(N)}\rangle:
		\\&&
		|a^{(1)}\rangle\in\mathcal C^{(1)}\wedge\ldots\wedge|a^{(N)}\rangle\in\mathcal C^{(N)}\}.
	\end{eqnarray}
	Especially, all states in $C^{(1,\ldots,N)}$ are fully separable.
	Using the convex hull of multipartite classical pure states as it was done in the bipartite case, we can extend the notion of classically correlated, multipartite states to mixed ones.
	Thus, we also constructed a concept of multipartite quantum correlations from the $N$ local definitions of classical states $\mathcal C^{(j)}$.

	As two examples, let us mention two tripartite quantum correlated states.
	They are
	\begin{eqnarray*}
		|\phi\rangle=\mathcal N(|a_1^{(1)}\rangle\otimes|a_1^{(2)}\rangle\otimes|a_1^{(3)}\rangle{+}|a_2^{(1)}\rangle\otimes|a_2^{(2)}\rangle\otimes|a_2^{(3)}\rangle)
	\end{eqnarray*}
	and
	\begin{eqnarray*}
		|\psi\rangle&=&\mathcal N(|a_1^{(1)}\rangle\otimes|a_1^{(2)}\rangle\otimes|a_2^{(3)}\rangle{+}|a_1^{(1)}\rangle\otimes|a_2^{(2)}\rangle\otimes|a_1^{(3)}\rangle
		\\&&{+}|a_2^{(1)}\rangle\otimes|a_1^{(2)}\rangle\otimes|a_1^{(3)}\rangle),
	\end{eqnarray*}
	with linearly independent $|a^{(j)}_r\rangle\in\mathcal C^{(j)}$ for $j=1,2,3$ and $r=1,2$.
	The state $|\phi\rangle$ corresponds to a GHZ state and $|\psi\rangle$ resembles a W state which represent inequivalent classes of full entanglement \cite{DVC00}.
	In addition, a partially entangled (or partially separable) state is $|\chi\rangle=|\hat a_1^{(1)}\rangle\otimes[|a_1^{(2)}\rangle\otimes|a_2^{(3)}\rangle+|a_2^{(2)}\rangle\otimes|a_1^{(3)}\rangle]$.

	As done for the bipartite case, we could now characterize the quantum correlations in more detail for such states, which would also require to specify the sets $\mathcal C^{(j)}$.
	Moreover, the quantification of the resulting nonclassical correlations can be performed as introduced in general in \cite{SV15}.
	There, the importance of quantum superpositions for constructing universal quantumness measures was also highlighted.

\subsection{Towards temporal correlations}
	Much more sophisticated is a generalization to include multitime correlations.
	For example, already the construction of multitime density operators is a complex problem \cite{APTV09}.
	Moreover, the question of temporal correlations can be closely related to a time evolution that has no counterpart in classical physics \cite{LG85,BKMPP15}.
	For mixed states this would mean that a description of a multitime system in terms of a stochastic process cannot be done.
	In a simplified manner, one could argue that the assumption of locality (or of local-hidden-variables) in the Einstein-Podolski-Rosen paradox \cite{EPR35} is replaced with the assumption of (classical) causality.

	As mentioned above, the time evolution of the system plays an important role for defining nonclassical temporal correlations.
	In particular, time-dependent Hamilton operators require a so-called time-ordering when formulating the solution of the equations of motion (\textit{e.g.} Schr\"odinger or Heisenberg equation).
	This is needed to define a proper unitary time evolution operator and it can have a major influence on the evolution of a system \cite{CBMS13,QS15,KSV16}.

	A concept of quantumness for which the multitime generalization has been successfully done is the harmonic oscillator system \cite{V08}.
	In this a system, the space-time dependent quantum correlations can be uncovered based on a $P$ functional, which generalizes the Glauber-Sudarshan function by including time-ordering effects.
	One example of its application is the interpretation of the prominent photon antibunching experiment \cite{KDM77}. 
	That is, the inability to interpret the $P$ functional, which describes the emission of photons, in terms of classical stochastic process.
	Recently, we have been also able to formulate a regularization approach to filter the $P$ functional \cite{KVS17}.

	In the future, it would be important to generalize the approach for temporal quantum correlations in harmonic oscillator systems to other systems and notions of nonclassicality.
	From such a technique, our method to construct multipartite quantum correlations from local ones could be generalized by including time-ordering effects to define multitime quantum correlations.
	Eventually, this could also result in a definition of temporal entanglement as the emerging quantum correlation that exists for any individual notion of nonclassicality. 
	However, this requires further research.

\section{Summary and conclusions}\label{sec:Summary}
	In this work, we analyzed quantum correlations in composite quantum systems.
	Especially, a bipartite state was defined and analytically characterized by combining one continuous-variable harmonic oscillator and one discrete-variable qubit subsystem.
	We could combine these seemingly incompatible concepts of nonclassicality of the individual parts and establish the concept of nonclassical correlations between the different degrees of freedom.
	This was achieved by requiring that a classically correlated system can be described in terms of uncorrelated, pure state and classical statistical mixing only.

	Some examples of pure nonclassical states have been studied.
	We showed the distinct differences of local and global quantum superpositions.
	The local superposition of classical states resulted in the nonclassicality of the subsystems, whereas globally superimposed states introduced nonclassical correlations between the two parts.

	A quantum state representation of the hybrid system under study was discussed and the classical states in this representation have been identified.
	This was based on a matrix quasiprobability distribution.
	We could distinguish between two classes of nonclassical states.
	Those are stares which can be written as pseudo-mixtures of classical ones and those which cannot be expanded in such a form at all.
	Moreover, a regularization approach was considered to resolve singularities within the matrix quasiprobability.

	Restricting ourselves to specific observables, we analyzed nonclassical correlations in connection to the joint and conditional probability distributions of the measurement outcomes.
	It was demonstrated that joint and conditional variances of the same observable and the same state state can be sensitive to different forms of nonclassicality.
	Let us emphasize that the conditional nonclassical correlations are connected to the remote manipulation of one subsystem by performing a measurement on the other.

	Finally, we also addressed more general forms of nonclassical correlations.
	We argued that entanglement is a quantum correlation which has some form of universal character.
	In general, entanglement emerges as one kind of quantum correlations among others.
	Yet, the remarkable fact is that it always emerges, which highlights the general importance of entanglement.
	We also outlined the generalization of our techniques to multipartite systems that consists of an arbitrary number of degrees of freedom and the possibility to include temporal correlations.

\ack
	This work was supported by the Deutsche Forschungsgemeinschaft through SFB 652, B12.
	The project leading to this application has received funding from the European Union's Horizon 2020 research and innovation programme under grant agreement No. 665148.

\end{document}